\documentclass[a4paper]{article}

\usepackage{INTERSPEECH2020}
\usepackage{balance,amsmath,graphicx,cite,multirow}

\title{Speaker-Utterance Dual Attention for Speaker and Utterance Verification}
\name{Tianchi Liu$^1$, Rohan Kumar Das$^2$, Maulik Madhavi$^2$, Shengmei Shen$^1$, Haizhou Li$^2$}
\address{
  $^1$Pensees Pte Ltd, Singapore\\
  $^2$Department of Electrical and Computer Engineering,
National University of Singapore, Singapore}
\email{\{liutianchi, jane.shen\}@pensees.ai, \{rohankd, maulik.madhavi, haizhou.li\}@nus.edu.sg}

\begin{document}

\maketitle
\begin{abstract}

In this paper, we study a novel technique that exploits the interaction between speaker traits and linguistic content to improve both speaker verification and utterance verification performance. We implement an idea of speaker-utterance dual attention (SUDA) in a unified neural network. The dual attention refers to an attention mechanism for the two tasks of speaker and utterance verification. The proposed SUDA features an attention mask mechanism to learn the interaction between the speaker and utterance information streams. This helps to focus only on the required information for respective task by masking the irrelevant counterparts. The studies conducted on RSR2015 corpus confirm that the proposed SUDA outperforms the framework without attention mask as well as several competitive systems for both speaker and utterance verification.

\end{abstract}

\vspace{2mm}

\section{Introduction}

Speaker verification (SV) aims to verify the claimed identity of a person using given speech~\cite{Campbell_sv}. Its implementation is broadly categorized into text-dependent and text-independent based on the spoken contents used for enrollment and testing~\cite{Campbell_sv}. The former deals with use of fixed short phases, while the latter doesn't put any constraints on the speech content. A text-independent system generally requires more training and test data~\cite{rkd_thesis,rkd_spcom_2016} than a text-dependent one~\cite{IET_biometric,IETE_rkd} to maintain  the same  level of accuracy. Therefore, text-dependent system is preferable in many real-world applications where user's cooperation is possible.

The research on text-dependent SV has evolved a lot from traditional dynamic time warping based template matching method to deep learning recently. For benchmarking of technology process, standard speech corpora like RSR2015 and RedDots are designed~\cite{rsr, Lee2015}. It is found that the modeling techniques such as hierarchical multi-layer acoustic model (HiLAM)~\cite{rsr}, unsupervised hidden Markov model (HMM)-universal background model (UBM)~\cite{Sarkar_is2016} i-vector/HMM~\cite{Zeinali_is2016} and j-vector~\cite{ChenQY15} benefit from temporal information in speech. Further, deep learning techniques have greatly improved the ability of speaker characterization~\cite{Liu_speechCom15,ShiLWL17,Heo2017,Dey_is2018,IS_ShiLLL18a,jvector_odysey2018}.

In human SV, we would like to have test samples as parallel to the unknown as possible to reduce the variability to a minimum due to language content. Text-dependent SV allows us to do just like that. 
Various studies show text-dependent SV is considered for performing two tasks, an SV as the main task, and an utterance verification as the subtask, where the two tasks are optimized separately or jointly in order to improve the SV objective. For example, the phonetic posteriorgrams derived using Gaussian mixture model (GMM) and deep neural network (DNN) frameworks are utilized to capture lexical information for text-dependent SV~\cite{rkd_is_2015,Dey_icassp_2016}. One shows that DNN based speaker embedding benefits from lexical content information~\cite{dey_icassp18}, others suggest that lexical information can be used in different ways to compensate the SV scores for performance gains~\cite{IS2014scheffer,Dey_content_is2017,APSIPA_uttcomp}.

Prior studies have underscored the importance of content modeling in text-dependent SV. While utterance verification has been well studied as part of speech recognition~\cite{uttver_jasa,uttver_TASLP}, it has not been given sufficient attention in the context of SV. Some consider text-dependent SV as a combination of two independent systems, namely SV and utterance verification~\cite{Kinnunen_is2016,Zeinali2018}. In our previous work~\cite{SUV_is2019}, text-dependent SV is formulated in a unified speaker-utterance verification (SUV) system as a multi-task learning  implementation. This is inspired by human cognitive process where we interpret and decode speaker traits and linguistic content in a corroborative manner \cite{IEEETran_jointTangLWV17,ISKumarYG18}. For example, by paying special attention to particular sounds while knowing the linguistic content information, we verify the voice of a speaker; on the other hand, if we are familiar with a speaker, we tend to recognize his/her voice in a better way.

In the unified SUV system, we used a shared long short term memory (LSTM) network and two independent LSTM output layers, one for speaker identity and another for utterance identity~\cite{SUV_is2019}. While the previously proposed unified SUV framework is effective, the interaction between the two output layers is not explored. We believe that both speaker and utterance verification can benefit from each other by exploring the temporal interaction between them. This is also motivated by successful explorations in text-dependent SV that suggest the benefit of compensating lexical information~\cite{IS2014scheffer,Dey_content_is2017,APSIPA_uttcomp}. Further, various attention models~\cite{zhang2016end,rahman2018attention,zhu2018self} project their possibility to focus on specific compensation or masking related to each task.

In this work, we propose an speaker-utterance dual attention, SUDA in short for performing both speaker and utterance verification. The attention mechanism for compensating irrelevant information for both tasks is derived by using masking operation. The masking for attention is estimated frame-by-frame, the attention mechanism establishes the temporal association between the speaker trait stream and the utterance content stream of LSTM output. In addition, we note that, as the attention mechanism is applied to both the branches (speaker and utterance) in the framework, it is referred to as dual attention. The studies in this work are conducted on RSR2015 corpus~\cite{rsr}. The contribution of this work lies in the use of a speaker-utterance dual attention in a single framework for performing both speaker and utterance verification.

The rest of the work is organized as follows. Section~\ref{secii} describes the proposed speaker-utterance dual attention mechanism for speaker and utterance verification. The experiments are detailed in Section~\ref{seciii}, followed by reporting of their results and analysis in Section~\ref{seciv}. The paper is finally concluded in Section~\ref{conc}.

\begin{figure*}[t!]
  \centering
  \includegraphics[width=\linewidth]{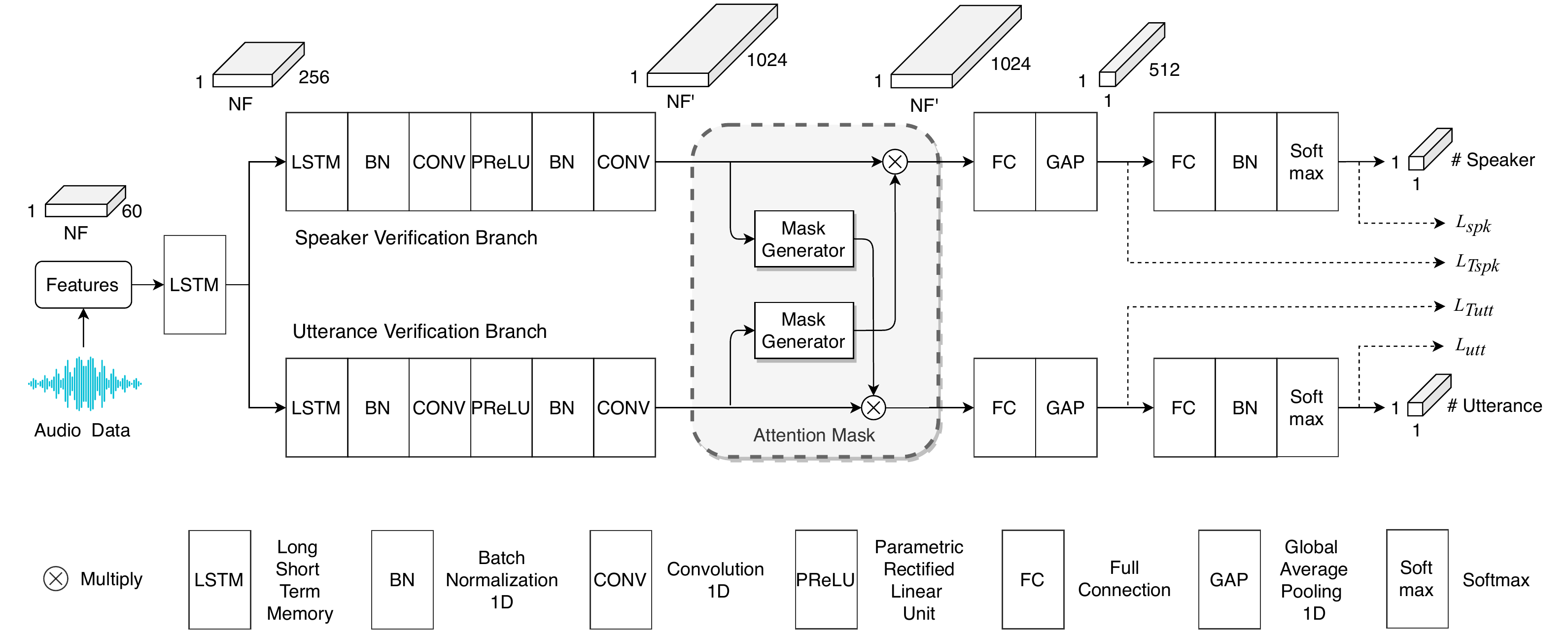}
  \caption{Block diagram of proposed SUDA for speaker and utterance verification. It consists of a shared LSTM, two LSTM output layer, which interact each other through the attention mask network. The NF and NF' are number of frames before and after convolution. }
  \label{fig:structure}
\end{figure*}

\section{Speaker-Utterance Dual Attention}
\label{secii}

This section describes the proposed SUDA for speaker and utterance verification. As shown in Figure~\ref{fig:structure}, the features extracted from raw audio data are fed into the LSTM based recurrent network to characterize the temporal dynamics. This system is an extension to our earlier work of unified SUV framework~\cite{SUV_is2019}. 

As presented in our earlier study \cite{SUV_is2019}, the first shared layer is common for speaker and utterance verification branches. Then, the hidden representation from the first shared layer is passed to the next two LSTM networks that focus on extracting valid information for each of the two sub-tasks, namely, speaker and utterance verification. We now discuss the improvisations introduced in this work using dual attention by masking.

\subsection{Related Studies}

Masking has been found to be effective to separate multiple sound sources from an audio mixture~\cite{WangC18a}.  A more recent approach regards speech separation problem as a supervised learning that aims to discriminate different patterns such as, speech, speakers, and background noise, which are learned from training data \cite{WangC18a}. 
Masking operation can be viewed as an attention mechanism, where the every single features from one branch are masked to attend the specific features from another branch. Attention based models have been applied successfully to various tasks. In computer vision research, there are three main strategies for attention: spatial attention, channel attention and mixed attention~\cite{jaderberg2015spatial, hu2018squeeze, wang2017residual}. The spatial attention is used to emphasis the area that one is interested, while the channel attention is mainly for features recalibration in convoluational neural networks, such as the `Squeeze-and-Excitation' block proposed in \cite{hu2018squeeze}. Again, the authors of \cite{zhang2016end} proposed an end-to-end framework with an attention model to combine the frame-level features, which acts as an alignment operation for SV studies. The attention model designed as a part of the speaker embedding network, is used to calculate the weighted mean of the frame-level feature maps to derive speaker embedding with a better discriminative speaker characteristics \cite{rahman2018attention,zhu2018self}.

\subsection{Attention Mask}
In the proposed SUDA, the convolution layers are used to extract the feature maps from the hidden representation obtained after LSTM as shown in Figure~\ref{fig:structure}. It has the objective to map the hidden representation to higher dimensional space to compensate the mutual information by attention masks. The dynamic masks can learn the information from feature maps obtained from the convolutional layers. We use sigmoid non-linearity to limit the activation from the mask. The dynamic masking is performed by learning the parameters from feature maps and the activating sigmoid function~\cite{mask_Xei}. The masking is formulated as:
\begin{equation}
  {\text{mask}}_s = 1 - {\text{Sigmoid}}(fm_u)
  \label{mask_spk}
\end{equation}
\begin{equation}
  {\text{mask}}_u = 1 - {\text{Sigmoid}}(fm_s)
  \label{mask_utt}
\end{equation}
where the $fm_{u}$ and $fm_{s}$ represent the feature maps from utterance verification and SV branch, respectively, while the ${\text{mask}}_{u}$ and ${\text{mask}}_{s}$ indicate the dynamic masks for the corresponding branch. These masks are then multiplied by the corresponding stream of feature maps to suppress irrelevant information in the respective stream of feature maps. The parameters in the masks are not fixed during the training or inference phase. They are derived according to the input audio data.

 \begin{table*}[t]
 \caption{Performance in EER (\%) for proposed SUDA in comparison to existing systems on RSR2015 Part I corpus.}

  \label{tab:table1-RSR2015PartI}
  \centering
  \begin{tabular}{c|ccc|ccc|ccc|ccc}
  
    \toprule
    \multirow{3}{*}{System} & \multicolumn{6}{c|}{Male} & \multicolumn{6}{c}{Female}\cr\cline{2-13}&
    \multicolumn{3}{c|}{Development} & \multicolumn{3}{c|}{Evaluation} &\multicolumn{3}{c|}{Development} & \multicolumn{3}{c}{Evaluation}\cr\cline{2-13} &
   TW&IC&IW&TW&IC&IW&TW&IC&IW&TW&IC&IW\cr\cline{1-13}
    
    \midrule
    i-vector \cite{rsr}     & 2.870 & 5.950 & 0.740 & 1.950 & 4.030 & 0.320 & 3.050 & 7.870 & 0.940 & 1.910 & 6.610 & 0.750\\
    HiLAM \cite{rsr}        & 1.660 & 3.690 & 0.490 & 0.820 & 2.470 & 0.190 & 1.770 & 3.240 & 0.450 & 0.610 & 2.960 & 0.140\\
    Joint-spk-utt \cite{SUV_is2019} & 5.565 & 1.981 & 1.792 & 5.125 & 2.079 & 0.888 & 5.179 & 1.699 & 0.831 & 3.110 & 1.453 & 0.499\\
    Unified SUV  \cite{SUV_is2019}  & 0.470 & 1.590 & 0.101 & 0.293 & 1.757 & 0.039 & 1.176 & 4.323 & 0.178 & 0.375 & 2.009 & 0.068\\
    Utt-comp \cite{APSIPA_uttcomp} &   -    & 1.460 &    -   &     -  & 0.960 &   -    &    -   & 1.640 &   -    &    -   & 0.730 &    -  \\               
    Utt-comp-uf \cite{APSIPA_uttcomp}  & - & 1.460 & - &   -   & 0.960 &  -     &      - & 1.460 &   -    &-      &\textbf{0.720}&-\\
      \cline{1-13}
    \bottomrule
 
    mod-SUV   & \textbf{0.202}  & 0.952  & 0.034  & 0.107 & 1.093 &  0.020 & 0.475 & 2.055 & 0.071  & 0.182 & 1.373 & 0.034\\
    {\bf Proposed: SUDA}     & \textbf{0.202} & \textbf{0.728} &  \textbf{0.022} &  \textbf{0.068} &  \textbf{0.722} &  \textbf{0.010} & \textbf{0.297} & \textbf{1.449} & \textbf{0.024} & \textbf{0.125} & 0.863 & \textbf{0.023}\\
    \cline{1-13} 
  \end{tabular}

\end{table*}

The operations performed so far produce the framewise representations. In order to pool the information across the utterance, global average pooling (GAP) is performed. Next, the fully connected (FC) layers are used to perform both speaker and utterance verification tasks.

In this work, the attention mask is applied on the every feature map through sigmoid operation than the conventional softmax operation. Further, the weights of our attention masks are obtained from the branch of another task. We note that the weights of the attention mask are tuned for different utterances and speaker-specific information.

\section{Experiments}
\label{seciii}

We now discuss the database and experimental setup for the studies in the following subsections.

\subsection{Database}
\label{subsec3-1}

The RSR2015 corpus is used for the studies in this work~\cite{rsr}. It contains 300 speakers data from 143 female and 157 male speakers. Further, the corpus is divided into three different parts based on the nature of the fixed phrases. The Part I includes 30 fixed phrase utterances of 3-4 seconds duration, whereas the Part II has 30 fixed short commands of 1-2 seconds duration. Similarly, the Part III contains the random digit based five or ten digit sequences. Again, there are 9 sessions for each phrase from all the speakers. Out of those, the first, fourth and seventh sessions are used for speaker enrollment and the remaining for testing as per the RSR2015 evaluation protocol~\cite{rsr}.

The RSR2015 corpus has three subsets, which are background, development and evaluation set to evaluate the system performance~\cite{rsr}.

The test trials are grouped under four categories based on the test speaker and phrase labels, which are {\textit {Target Correct}} (TC), {\textit {Impostor Correct}} (IC), {\textit {Target Wrong}} (TW) and {\textit {Impostor Wrong}} (IW). They further constitute three test conditions, where each of those conditions consider TC as target trials and the remaining three categories as non-target trials, respectively. The performance is reported in terms of equal error rate (EER). In this work, we consider Part I and Part II of RSR2015 as they are suitable for both speaker and utterance verification.

\subsection{Experimental Setup}
\label{subsec3-2}

The speech utterances are processed with 20 ms frame size and 10 ms shift to extract 60-dimensional (20-base + 20-$\Delta$ + 20-$\Delta\Delta$) mel frequency cepstral coefficient (MFCC) features using KALDI\footnotemark[1] toolkit. The extracted features are normalized by cepstral mean and variance normalization using utterance level mean and variance statistics.

\footnotetext[1]{http://kaldi-asr.org/}
\footnotetext[2]{https://pytorch.org/}

In addition, we apply the feature level triplet loss to further increase the intra-class distance and reduce inter-class distance. It is applied on the 512-dimensional feature vector at the step after 1D global average pooling as shown in Figure~\ref{fig:structure}. We calculate triplet loss in both the branches, i.e., SV and utterance verification. In addition, the negative and positive samples in the training batch for each iteration is searched. The batch size of each iteration is set to be 128 for all the experiments and these samples are chosen randomly.
The loss of proposed framework is calculated by following formula:
\begin{equation}
  L_{total} = L_{Tspk} + L_{Tutt} + L_{spk} + L_{utt}
  \label{loss}
\end{equation}
where the $L_{Tspk}$ and $L_{Tutt}$ are triplet loss, while $L_{spk}$ and $L_{utt}$ are negative log likelihood loss. The subscripts $spk$ and $utt$ represent the SV and utterance verification branches in SUDA, respectively.

\begin{table*}[t]
  \caption{Performance in EER (\%) for proposed SUDA in comparison to existing systems on RSR2015 Part II corpus.}

  \label{tab:table2-RSR2015PartII}
  \centering
  \begin{tabular}{c|ccc|ccc|ccc|ccc}
  
    \toprule
    \multirow{3}{*}{System} & \multicolumn{6}{c|}{Male} & \multicolumn{6}{c}{Female}\cr\cline{2-13}&
    \multicolumn{3}{c|}{Development} & \multicolumn{3}{c|}{Evaluation} &\multicolumn{3}{c|}{Development} & \multicolumn{3}{c}{Evaluation}\cr\cline{2-13} &
   TW&IC&IW&TW&IC&IW&TW&IC&IW&TW&IC&IW\cr\cline{1-13}
    
    \midrule
    i-vector \cite{rsr}     & 5.410 &13.750 & 2.500 & 4.390 &11.260 & 1.810 & 6.940 &12.730 & 2.860 & 5.160 &15.270 & 3.050\\
    HiLAM \cite{rsr}        & 6.140 &10.580 & 3.030 & 4.420 & 8.380 & 1.710 & 4.620 & 6.660 & 1.290 & 3.710 & 7.950 & 1.450\\
    Joint-spk-utt   &10.804 & 4.096 & 2.715 & 9.929 & 4.190 & 2.286 &10.220 & 3.482 & 2.179 & 7.797 & 3.382 & 1.816\\
    Utt-comp \cite{APSIPA_uttcomp} &    -   & 4.160 &    -   &    -   & 3.610 &   -    &     -  & 4.030 &  -     &     -  & 2.850 &   -   \\                           
    Utt-comp-uf \cite{APSIPA_uttcomp} & - & 4.160 & -  &   -    & 3.610 &   -    &   -    & 3.860 &    -   &   -    & 2.790 &  -    \\
    \cline{1-13}
    
    \bottomrule

    mod-SUV   & 1.394 & 3.757 & 0.279 & 1.015 & 3.591 &  0.176 & 1.833 & 4.862 &  \textbf{0.272}  & 0.851 & 3.563 & 0.102\\
    {\bf Proposed: SUDA}     & \textbf{1.382} & \textbf{2.698} & \textbf{0.245} &  \textbf{0.878} & \textbf{2.400} & \textbf{0.127} & \textbf{1.360} &  \textbf{3.359} & 0.296 & \textbf{0.624} & \textbf{2.020} & \textbf{0.057}\\
    \cline{1-13} 
  \end{tabular}
  \vspace{-2mm}
\end{table*}

The learning rate, optimizer, LSTM's hidden layer and scoring follow the same configurations as that in our previous work of unified SUV~\cite{SUV_is2019}. We adopt PyTorch\footnotemark[2] toolkit for the implementation. We fixed seeds empirically to 2020 in our studies. In contrast to the previous unified SUV, we add 1D convolution layer to the network, where kernel size of convolution layer is 5, while the padding and the stride are 0 and 1, respectively. The activation function between convolutional layers is parametric rectified linear unit (PReLU) \cite{he2015delving}. Further, we note that in order to observe the impact of attention mask, we also conduct the experiments without attention based on masking operation block. We refer to the system with this setup as mod-SUV for comparing to previous unified SUV~\cite{SUV_is2019} in this work.

\section{Results and Discussions}
\label{seciv}

\begin{table}[t]
  \caption{Performance in EER (\%) of different systems on the evaluation set of RSR2015 Part I. Here, j-vector: j-vector with cosine similarity,  Joint Bayesian: j-vector system with joint Bayesian model, J2: joint training of j-vector extractor and joint Bayesian and the Siamese network for j-vector extractor and the joint Bayesian as a back-end, J3: joint training of j-vector extractor and joint Bayesian, and use the Siamese network output for verification, RACNN-LSTM: raw audio convolutional neural network with LSTM, i-vector + s-vector: the system concatenating i-vector and s-vector directly, i-s-vector: the system concatenating the last-step hidden output of s-vector and corresponding i-vector (s-vector extracted either from LSTM or Bidirectional LSTM (BLSTM)).}

  \label{tab:table3-RSR2015Part I evaluation set - combine gender}
  \centering
  \begin{tabular}{c|ccc}
    \toprule
    System & TW&IC&IW \\

    \midrule
    \cline{1-4}
    j-vector\cite{ShiLWL17}    & 3.14 & 7.86 & 0.95 \\
    Joint Bayesian \cite{IS_ShiLLL18a}  & 0.03 & 3.61 & 0.02 \\
    J2 \cite{IS_ShiLLL18a}   & \textbf{0.02} & 2.81 & 0.02 \\
    J3 \cite{IS_ShiLLL18a}   &\textbf{0.02}& 2.42 &0.02\\   
    RACNN-LSTM \cite{jung2018complete}         & - & 3.63 & -\\  
    Unified SUV  \cite{SUV_is2019}  &0.46& 2.41 &0.06\\
    i-vector + s-vector \cite{wang2017does} &0.28 &1.13 &0.03 \\
    i-s-vector (LSTM) \cite{wang2017does} &0.17&1.98&0.03\\
    i-s-vector (BLSTM) \cite{wang2017does} &0.11&1.72&0.02\\
    \cline{1-4}
    \bottomrule

    mod-SUV        & 0.15 & 1.14 &  0.02\\
    {\bf Proposed: SUDA}          & 0.13 & \textbf{0.62} & \textbf{0.01}\\
    \cline{1-4}
  \end{tabular}
  \vspace{-2mm}
\end{table}

We consider HiLAM and i-vectors as two basic common reference systems for the studies~\cite{rsr}. Further, as the work advocates on compensation network, we consider joint speaker utterance (joint-spk-utt)~\cite{Wang_is2016} and utterance compensation (utt-comp)~\cite{APSIPA_uttcomp} frameworks for comparison. We note that joint speaker utterance models speaker and utterance information jointly~\cite{Wang_is2016}, whereas the utterance compensation framework compensates the utterance information after jointly modeling speaker and utterance characteristics~\cite{APSIPA_uttcomp}. Further, the utterance compensation framework has another variant with utterance factor (utt-comp-uf). We note that all these works used for comparison targets for only SV studies. The unified SUV proposed in our previous work for performing both speaker and utterance verification is also used as reference system~\cite{SUV_is2019}.

Table~\ref{tab:table1-RSR2015PartI} shows the performance comparison of our proposed SUDA to the systems discussed above on Part I of RSR2015 corpus. The performance of systems compared are quoted from the previously published results. We find that HiLAM system performs better than the i-vectors systems due to use of temporal knowledge. Further, the joint speaker utterance model outperforms HiLAM system for IC test trial condition, while it performs poorly in TW and IW trial conditions related to utterance verification. Compared to these two systems, our previous work, unified SUV takes advantage of LSTM to capture temporal dynamics, greatly reduces the error rate of TW and IW trial conditions. In the utterance compensation system, we observe compensating utterance information leads to a good performance in IC trial condition that further improves in addition of the utterance factor \cite{APSIPA_uttcomp}. However, it did not explore on compensating speaker information for utterance verification unlike this paper, hence, the results of TW and IW trials are not investigated in \cite{APSIPA_uttcomp}.

In this work, as mentioned in Section \ref{subsec3-2}, mod-SUV is a modified framework of unified SUV. We can observe from Table \ref{tab:table1-RSR2015PartI} that mod-SUV has significantly improved SV performance in all three test trial conditions over the existing unified SUV framework. Further, our proposed SUDA, which focuses on the required speaker and utterance information by imposing dual attention, outperforms most of the other systems for all three test trials conditions except, IC test trial condition of evaluation female set.

Table \ref{tab:table2-RSR2015PartII} reports the performance of various systems on Part II of RSR2015 database. The performance trend of various systems remains similar to that observed in case of Part I of RSR2015 database. The proposed SUDA again outperforms all other systems, showing effectiveness of attention based dual compensation in various trial conditions.

We now compare our proposed SUDA framework with other deep learning systems. We combine both male and female data of RSR2015 Part I to match with the evaluation protocol followed in~\cite{ShiLWL17,IS_ShiLLL18a,jung2018complete,wang2017does} for comparison with other research studies. We observe from Table \ref{tab:table3-RSR2015Part I evaluation set - combine gender} that the mod-SUV has a comparable performance to other systems in the IC trial condition. Further, with the use of attention masks, the proposed SUDA achieves a significant improvement on the IC trial condition, and at the same time improves performance for IW and TW trial conditions.

Finally, as discussed in our earlier work \cite{SUV_is2019}, we can adjust and tune the scores from speaker and utterance verification branch to show a security trade-off during scoring. The studies related to this reported in Table \ref{tab:table4-RSR2015Part I evaluation set - SV and UV} show that the proposed SUDA outperforms previous unified SUV and the current mod-SUV, that highlights the gain provided by dual attention mechanism.

\begin{table}[t]
  \caption{Performance in EER (\%) for SV and utterance verification (UV) on Part I of RSR2015 evaluation set.}

  \label{tab:table4-RSR2015Part I evaluation set - SV and UV}
  \centering
  \begin{tabular}{c|cc|cc}
    \toprule
    \multirow{2}{*}{System} & \multicolumn{2}{c|}{Male} & \multicolumn{2}{c}{Female}\cr\cline{2-5}&
    SV&UV&SV&UV\cr  
    \midrule
     \cline{1-5}
                         
    Unified SUV  \cite{SUV_is2019}      &1.796& 0.021 &1.918&\textbf{0.011}\\
    \cline{1-5}
    \bottomrule

    mod-SUV  &  1.132 & 0.010 & 1.362 & \textbf{0.011}\\
    {\bf Proposed: SUDA}    &  \textbf{0.741} & \textbf{0.005} &\textbf{0.840} & \textbf{0.011}\\
    \hline
  \end{tabular}
  \vspace{-4mm}
\end{table}

\section{Conclusions}
\label{conc}

This work proposes a novel speaker-utterance dual attention (SUDA) mechanism for speaker and utterance verification. We used LSTM based models with two branches in a unified framework that consider attention masks for temporal interaction between speaker trait stream and utterance content stream that helps to suppress the irrelevant information for both tasks. The studies conducted on RSR2015 corpus reveal the importance of the proposed SUDA in comparison to existing approaches to work effectively for both speaker and utterance verification simultaneously. The framework also leverages a user for using it according to the security need of the intended application. The future work will focus on extending attention masking for prompted digit based SV.

\section{Acknowledgements}

This research is supported by Programmatic Grant No. A1687b0033 from the Singapore Government's Research, Innovation and Enterprise 2020 plan (Advanced Manufacturing and Engineering domain), and Human-Robot Interaction Phase 1 (Grant No. 192 25 00054) by the National Research Foundation, Prime Minister's Office, Singapore under the National Robotics Programme.

\balance
\bibliographystyle{IEEEtran}

\bibliography{Tianchi_paper_PostReview_v7}

\end{document}